\documentclass[12pt]{JHEP3}

\usepackage{amsmath}
\usepackage{longtable}

\newcommand{\be}{\begin{eqnarray}}
\newcommand{\ee}{\end{eqnarray}}

\newcommand{\bn}{\begin{enumerate}}
\newcommand{\en}{\end{enumerate}}

\def\det{{\rm det}}

\title{Evidence for
Aharony duality for orthogonal gauge groups }

\author{ Chiung Hwang $^{1}$, Kyung-Jae Park$^{1}$, Jaemo Park$^{1,2}$

\\

\\

$^1$Department of Physics, POSTECH, Pohang 790-784, Korea
\\
$^2$Postech Center for Theoretical Physics (PCTP), Postech, Pohang
  790-784, Korea

\\
\\
E-mail: \email{ilvhemos@postech.ac.kr, jaco@postech.ac.kr,
jaemo@postech.ac.kr} } 

\abstract{ We study the Aharony duality for three dimensional
$\mathcal N=2$ supersymmetric gauge theories for orthogonal gauge
groups with matters in vector representation. We provide the
evidence for the duality by working out the partition function on
$S^3$ and the superconformal index, which show perfect agreement. }

\begin{document}

\section{Introduction}

We have witnessed the tremendous progress in understanding the 3d
SCFTs recently. One of the key observations was put forward by J.
Schwarz that such theories could be described by Super Chern-Simons
matter theories(SCSM)\cite{Schwarz04}. This led to  rapid progress
in $AdS_4/CFT_3$ correspondence\cite{BL1, BL2, BL3, gus1, gus2,
GaiottoWitten, Hosomichi08,Aharony08, Hosomichi08a,
Imamura08,Gaiotto07} and the understanding the CFT associated with
M2 branes\cite{Hanany1, Hanany2, Hanany3, Martelli, Jafferis09,
Benini10}. Along with this development, there also have been
sophisticated tools developed to probe 3d SCFTs such as the
partition function on $S^3$\cite{KWY1, Kapustin11,Willett11} and the
superconformal index\cite{Kim09,Imamura11,HKPP11,Bashkirov}, which
gives the detailed information on the 3d SCFTs. Furthermore in this
setting 3d SCFTs do not have to be realized by SCSM type. In the
Yang-Mills type theories of 3d, the YM kinetic term is irrelevant in
the IR and one can take simply $g_{YM}\rightarrow \infty$ in the
partition function and the superconformal index in many examples.
With such sophisticated tools available one can understand the
various dualities in 3d far better than before. Such examples are
mirror symmetry and Seiberg-like dualities for SCSM
theories\cite{Giveon09, Niarchos, Koh11, Vartanov11}. Also one can
find the evidences that some YM type theories are flowing to SCSM
type SCFTs in the IR \cite{Koh11}.

In this paper, we are interested in the Aharony
duality\cite{Aharony97}. It was shown that Seiberg-like duality for
SCSM theories could be derived from Aharony duality. In the
available literatures this duality was discussed for U/Sp gauge
group where the index computation and the partition function give
impressive confirmation of the claimed
duality\cite{Willett11,Bashkirov}. Curiously the discussion on the
Aharony duality for orthogonal groups are lacking. Here we are
filling the gap by working out the partition function (numerically)
and the superconformal index to show that the duality works for
orthogonal groups with matters in the vector representation.

In  section 2, we propose the Aharony duality for orthogonal groups with matters in the vector representation
and first work out the partition function on $S^3$ numerically, which exhibits nice agreement between electric theory
and magnetic theory. Also we work out the $R$-charge of the various fields
by using the Z-minimization procedure for simple cases \cite{Jafferis10}. In section 3, we work out the superconformal index
and see the perfect matching in both sides. We also discuss chiral ring structures and enumerate
operators of lower dimensions appearing in the index. Then we discuss and conclude.
As this work is  near the final stage, we are aware of the work \cite{Benini11}, which
also discusses the Aharony duality for orthogonal groups.

\section{Partition function on $S^3$}
We consider the following electric and magnetic pair of $\mathcal N=2$ gauge
theory in 3-d.

\begin{itemize}
\item The electric theory
is the $\mathcal N=2$ $O(N_c)$ supersymmetric gauge  theory with
$N_f$ flavors of chiral multiplets $Q^a$ in the vector
representation with no superpotential.

\item The magnetic theory  is the $\mathcal N=2$ $O(N_f-N_c+2)$
supersymmetric gauge theory with $N_f$ flavors of chiral multiplets
$q_a$ in the vector representation as well as gauge singlet chiral
multiplets $M^{\{ab\}}$ and $Y$.  The magnetic theory has the
tree-level superpotential
\begin{equation}\label{superpotential}
W=M^{\{ab\}}q^i_a q^i_b+Yy,
\end{equation}
where $y$ is the monopole operator, parametrizing the Coulomb branch
of the magnetic theory.

\end{itemize}
  Here $Y$ is fundamental (non-composite)
field while $y$ is the monopole operator, which can be expressed in
terms of other fields. In fact $Y$ is generically mapped under the
duality to the monopole operator of the electric theory. We use the
same $Y$ to denote the monopole operator in the electric side.

The global symmetries and the corresponding charges of
the elementary fields and the monopole fields are listed in Table
\ref{charge}.
\begin{table}
\begin{center}
\begin{tabular}{|c|ccc|}
\hline
Fields & $U(1)_{R}$ & $U(1)_A$ & $SU(N_f)$\\
\hline
$Q^a$ & $r$ & 1 & $\mathbf N_f$\\
$Y$ & $N_f-N_c+2-N_f r$ & $-N_f$ & $\mathbf 1$\\
\hline
$q_a$ & $1-r$ & $-1$ & $\mathbf {\bar N_f}$\\
$M^{\{ab\}}$ & 2r & 2 & $\mathbf {N_f(N_f+1)/2}$\\
$Y$ & $N_f-N_c+2-N_f r$ & $-N_f$ & $\mathbf 1$\\
$y$ & $N_c-N_f+N_f r$ & $N_f$ & $\mathbf 1$\\
\hline
\end{tabular}
\caption{The global symmetry charges of the elementary fields and the monopole fields.}\label{charge}
\end{center}
\end{table}
In the table, $r$ denotes the $R$-charge of the fields $Q^a$ in the
IR limit, which is the same as its conformal dimension. In the UV
limit, $r=\frac{1}{2}$.  There are several cases worthy of mention.
For electric $O(1)$, there would be no monopole operator.
Nevertheless, when the magnetic theory admits monopole operator $y$,
we have to introduce the singlet operator $Y$ whose conformal
dimension and other quantum numbers are dictated by the
superpotential term $W=Yy+\cdots$. This additional singlet is
crucial to match the partition function and the superconformal index
as we will see later. On the other hand, if the magnetic group is
$O(1)$, while electric side has the monopole operator, we have to
introduce the singlet $Y$ in the magnetic side, whose conformal
dimension is the same as that of the monopole operator in the
electric side. In this case the superpotential term is
$W=M^{\{ab\}}q^i_a q^i_b +Y^2 {\det M}$. Its existence is motivated
by the similar reasoning to eq. (\ref{specialpotential}). This term
is also discussed in \cite{Aharony11}.

Also $O(2)$ is interesting. If we consider the $SO(2)$ gauge theory,
this has two independent monopole operators $Y_+, Y_-$ since it is
isomorphic to $U(1)$. However under nontrivial $Z_2$ action of
$O(2)$ $Y_+$ is mapped to $Y_-$ thus we have to consider $Z_2$
invariant combination $Y\equiv \frac{Y_++Y_-}{\sqrt{2}}$. Thus it is
crucial to consider $O(N)$ gauge theory instead of $SO(N)$ for
Aharony duality. This also removes the topological current $J=*dA$
for abelian theory with $A$ being the gauge potential, whose dual in
the nonabelian case is not clear.

The above Aharony duality can be motivated by considering the
Hanany-Witten setup with $D3-NS5_{\theta}-NS5_{-\theta}-O5$ and the
brane move passing through the infinite coupling where
$NS5_{\theta}$ and $NS5_{-\theta}$ branes are coincident
\cite{GiveonKutasov98}. Here $D3$  spans (0126), NS5 spans (012789)
and NS5' and O5 span (012345) and (012789) respectively.
$NS5_{\theta}$ denotes the rotated $NS5$ brane by $\theta$ in
(345)-(789) planes. We put the flavor D5 branes parallel to
$NS5_{\theta}$ to avoid the quartic superpotential. For orthogonal
case, we can consider D3 in the presence of $O5^+$ plane. Due to the
fact that  RR charge of $O5^+$ plane is the same as D5 brane,
starting from $O(N_c)$ gauge theory one ends up with $O(N_f-N_c+2)$
theory. By this way one can guess the above form of the Aharony
duality. And for symplectic case, one has to consider D3s with
$O5^-$ plane.

One can use the partition function on $S^3$ and the superconformal
index to give evidence for this conjecture. As a first test, one can
work out the partition function on $S^3$ for the theories on both
sides. One can also find the $R$-charges of $Q^a$ using the
Z-minimization \cite{Jafferis10}. For other discussions on the
partition function on $S^3$, see \cite{Vartanov11-2, Vartanov11-3}.

Supersymmetric localization gives the following expression for the
partition function of the ``electric'' theory \cite{KWY1,Jafferis10}:
\begin{eqnarray}
Z^{el,N_f}_{N_c}(\Delta_Q)=\frac{1}{|\mathcal W|}\int  \left(\prod_a
du^a\right) F_{N_c}(u) e^{N_f G_{N_c}(u,\Delta_Q)},
\end{eqnarray}
where $\Delta_Q$ denote the conformal dimension of $Q^a$ in the
electric side, the variables $u^a$ are real, the indices $a,b$ range
from $1$ to $\left[N_c/2\right]$, and $|\mathcal W|$ is the order of
the Weyl group $\mathcal W$. The expressions for functions $F_{N_c}$
and $G_{N_c}$ depend on whether $N_c$ is even or odd. If $N_c$ is
even, say, $N_c=2n$, we have
\begin{eqnarray}
&& F_{2n}(u)=\prod_{a<b} \left(4 \sinh(\pi (u_a-u_b))\sinh (\pi(u_a+u_b))\right)^2,\\
&& G_{2n}(u,\Delta_Q)=\sum_a \left(l(1-\Delta_Q+i u_a)+l(1-\Delta_Q-i u_a)\right).
\end{eqnarray}
Here the function $l(z)$ is given by
\begin{equation}\label{eq:lz}
l(z)=-z {\rm log} (1-e^{2\pi iz})+\frac{i}{2} \left(\pi
z^2+\frac{1}{\pi}{\rm Li}_2(e^{2\pi i z})\right)-\frac{i\pi}{12},
\end{equation}
which satisfies $dl(z)/dz=-\pi z \cot(\pi z)$.
If $N_c$ is odd, $N_c=2n+1$, we have
\begin{eqnarray}
&& F_{2n+1}(u)=\prod_c (2\sinh(\pi u_c))^2\prod_{a<b} \left(4 \sinh(\pi (u_a-u_b))\sinh (\pi(u_a+u_b))\right)^2,\\
&& G_{2n+1}(u,\Delta_Q)=l(1-\Delta_Q)+\sum_a \left(l(1-\Delta_Q+i u_a)+l(1-\Delta_Q-i u_a)\right).
\end{eqnarray}
The partition function of the ``magnetic'' theory is similar:
$$
Z^{mag,N_f}_{N_c'}(\Delta_q,\Delta_M,\Delta_Y)=\frac{1}{|\mathcal
W'|}e^{l(1-\Delta_Y)}e^{l(1-\Delta_M)N_f(N_f+1)/2} \int \left(\prod_a du^a\right)
F_{N_c'}(u) e^{N_f G_{N_c'}(u,\Delta_q)},
$$
where $N_c'=N_f-N_c+2$ and the indices $a,b$ now range from $1$ to
$[N_c'/2]$. The pre-factor is the contribution of the gauge singlet
chiral multiplets. As argued in \cite{Kapustin11, Willett11}, the
partition function should be the same for electric theory and
magnetic theory as a function of $\Delta_Q$. It's convenient to list
the free energy  $F=-\mathrm{log}|Z|$. We check the partition
function on $S^3$ for a few cases and find the agreement up to the
accuracy of $10^{-6}$. Some of the simple cases are listed in Table
\ref{free energy 1}, \ref{free energy 2}, \ref{free energy 3} .

\begin{table}
\begin{center}
\begin{tabular}{|c|cccccccc|}
\hline
$\Delta_Q$ & 0.275 & 0.3 & 0.325 & 0.35 & 0.375 & 0.4 & 0.425 & 0.45  \\
\hline
$-\mathrm{log}|Z|$ & 1.66573 & 1.75794 & 1.8338 & 1.88659 & 1.91895 & 1.93301 & 1.93056 & 1.91308 \\
\hline
\end{tabular}
\caption{Free energy $F=-\mathrm{log}|Z|$ for $O(2)^{el}_2$ and its dual $O(2)^{mag}_2$ theory.}\label{free energy 1}
\end{center}
\end{table}
\begin{table}
\begin{center}
\begin{tabular}{|c|cccccccc|}
\hline
$\Delta_Q$ & 0.275 & 0.3 & 0.325 & 0.35 & 0.375 & 0.4 & 0.425 & 0.45  \\
\hline
$-\mathrm{log}|Z|$ & 2.24168  & 2.43027 & 2.5767 & 2.68642 & 2.76383 & 2.81249 & 2.83538 & 2.83499 \\
\hline
\end{tabular}
\caption{Free energy $F=-\mathrm{log}|Z|$ for $O(2)^{el}_3$ and its dual $O(3)^{mag}_3$ theory.}\label{free energy 2}
\end{center}
\end{table}
\begin{table}
\begin{center}
\begin{tabular}{|c|cccccccc|}
\hline
$\Delta_Q$ & 0.275 & 0.3 & 0.325 & 0.35 & 0.375 & 0.4 & 0.425 & 0.45  \\
\hline
$-\mathrm{log}|Z|$ & 4.76985 & 5.05169 & 5.25322 & 5.38042 & 5.45165 & 5.43218 & 5.42046 & 5.33034 \\
\hline
\end{tabular}
\caption{Free energy $F=-\mathrm{log}|Z|$ for $O(3)^{el}_4$ and its dual $O(3)^{mag}_4$ theory.}\label{free energy 3}
\end{center}
\end{table}

We can obtain the conformal dimension of fields by  extremizing $\log|Z|$.
Generally, it is hard to find the conformal dimension that extremizing the partition function analytically but for the simple case of $N_f-N_c+2=1$, the dual magnetic theory becomes theory of $O(1)$ and it is relatively easy to find. The partition function of magnetic $O(1)$ theory is
\begin{eqnarray}
  \label{eq:z}
  Z_{N_c}^{mag,N_f}=Z_{N_f+1}^{mag,N_f}&=&e^{l(1-\Delta_Y)}e^{l(1-\Delta_M))N_f(N_f+1)/2}e^{N_fl(1-\Delta_q)}\nonumber\\
  &=&e^{l(N_f r)+l(1-2r)N_f(N_f+1)/2+N_f l(r)}
\end{eqnarray}
where $\Delta_Y=N_c'-N_f r$, $\Delta_M=2r$ and $\Delta_q=1-r$.
This expression is real positive, so it is effectively the same as extremizing its logarithm
\begin{eqnarray}
  \label{eq:logz}
  0&=&\frac{d\log Z}{d r}\nonumber\\
  &=&-\pi N_f^2 r \cot(\pi N_f r)+\pi N_f(N_f+1)(1-2r)\cot(\pi(1-2r))-\pi N_f r\cot(\pi
  r).
\end{eqnarray}
 We can find the analytic solution of \eqref{eq:logz} when $N_f=1$,
$r=\Delta_Q=1/3$. The magnetic side is $O(1)=Z_2$ gauge theory with
one flavor $q$, two singlets $Y, M$ with the superpotential
\begin{equation}
W=M (q^2+Y^2).
\end{equation}
Thus the R-charge of $q$ is $\frac{2}{3}$ and that of $Y$ is $\frac{2}{3}$. Note that would-be monopole
operator $y$ has conformal dimension $-\frac{4}{3}$  if coupled to $Y$ via the superpotential term $W=Yy+\cdots$.
This is in violation of the unitarity and is consistent with fact that for $O(1)$ theory we do not have
the monopole operator. Also $Z_2$ acts on $q$ by flipping its sign. Note that for both  electric and magnetic side,
the moduli space is parametrized by the gauge invariant operators $Y, M$.

Generically we can find the conformal dimension numerically, so when
$N_f=2$, $r=0.2697$ and when $N_f=3$, $r=0.4256$ and they are within
the unitarity bound   $\Delta_M=2\Delta_Q\geq 1/2$. We can see that
the $\Delta_M$ is irrational. These values we get from the above are
coincident with those of the corresponding electric theory obtained
numerically as listed at Table \ref{Conformal dimension}.
\begin{table}
\begin{center}
\begin{tabular}{|c|cccccccc|}
\hline
Theory & $O(2)_1$ & $O(2)_2$ & $O(2)_3$ & $O(2)_4$ & $O(3)_2$ & $O(3)_3$ & $O(3)_4$ & $O(3)_5$ \\
\hline
$\Delta_Q$ & 0.3333 & 0.4086 & 0.4370 & 0.4520 & 0.2697 & 0.3531 & 0.3923 & 0.4149 \\
\hline
\end{tabular}
\caption{Conformal dimensions for various ``electric''
theories.}\label{Conformal dimension}
\end{center}
\end{table}

Consider ``electric'' $O(2)_2$ theory as an example. Its dual
``magnetic'' theory is $O(2)_2$. The critical value of the dimension
is $\Delta_Q=0.4086$. And ``electric'' $O(2)_3$ case, the critical
value of $\Delta_Q=0.4370$. Note that the conformal dimension
$\Delta_Q$ become closer to $1/2$ as $N_c/N_f$ decreases. Since the
theory is more  weakly coupled in this limit, this is an expected
result. For some  other cases, we list the conformal dimension of
$Q$ in Table \ref{Conformal dimension}.

\section{Superconformal index}
We consider the superconformal index for 3-d $\mathcal{N}=2$
superconformal field theory (SCFT). The bosonic subgroup of the 3-d
$\mathcal{N}=2$ superconformal algebra is $SO(2,3) \times SO(2) $.
There are three Cartan elements denoted by $\epsilon, j_3$ and $R$
which come from three factors $SO(2)_\epsilon \times
SO(3)_{j_3}\times SO(2)_R $ in the bosonic subalgebra.  One can
define the superconformal index for 3-d $\mathcal{N}=2$ SCFT as
follows \cite{Bhattacharya09},
\begin{equation}
I=Tr (-1)^F exp
 (-\beta'\{Q, S\})
x^{\epsilon+j_3}\prod_j y_j^{F_j}
\label{def:index}
\end{equation}
where $Q$ is a special  supercharge with quantum numbers $\epsilon =
\frac{1}2, j_3 = -\frac{1}{2}$ and $R=1$ and $S= Q^\dagger$. They
satisfy the following anti-commutation relation:
\begin{equation}
 \{Q, S\}=\epsilon-R-j_3 : = \Delta.
\end{equation}
In the index formula, the trace is taken over gauge-invariant local
operators in the SCFT defined on $\mathbb{R}^{1,2}$ or over states
in the SCFT on $\mathbb{R}\times S^2$. As is usual for Witten index
, only BPS states satisfying the bound $\Delta =0 $ contributes to
the index and the index is independent of $\beta'$. If we have
additional conserved charges commuting with chosen supercharges
($Q,S$), we can turn on the associated chemical potentials and the
index counts the number of BPS states with the specified quantum
number of the conserved charges denoted by $F_j$ in eq.
(\ref{def:index}).

The superconformal index is exactly calculable using localization
technique \cite{Kim09,Imamura11}.  Following their works, the
superconformal index can be written in the following form,
\begin{equation}\label{formula:index}
I(x)=\sum_{m} \int da\, \frac{1}{|\mathcal W|}
e^{-S^{(0)}_{CS}}e^{ib_0(a)} y_{j}^{q_{0j}}
x^{\epsilon_0}\exp\left[\sum^\infty_{n=1}\frac{1}{n}f_{tot}(e^{ina},
y_{j}^n,x^n)\right].
\end{equation}
To take trace over Hilbert-space on $S^2$, we impose proper periodic
boundary conditions on time direction $\mathbb{R}$. As a result, the
base manifold become $S^1\times S^2$. For saddle points in
localization procedure, we need to turn on monopole fluxes on $S^2$
and holonomy along $S^1$. These configurations of the gauge fields
are denoted by  $\{m\}$ and $\{ a \}$ collectively. Both variables
take values in the Cartan subalgebra of $G$. $S_0$ denote the
classical action for the (monopole+holnomy) configuration on
$S^1\times S^2$. $\epsilon_0$ is called the Casimir energy.
 If we consider the theory without Chern-Simons (CS) term, there would be
 no contribution from $S_0$. Also if we consider non-chiral theories, we have $b_0(a)=0$.
 In \eqref{formula:index}, $\sum_{m}$ is
over all integral magnetic monopoles charges,
$f_{tot}=f_{chiral}+f_{vector}$ and $|\mathcal W| = (\text{the order
of the Weyl group})$.  Each component in \eqref{formula:index} is given by
\begin{eqnarray}
&&b_0(a)=-\frac{1}{2}\sum_\Phi\sum_{\rho\in R_\Phi}|\rho(m)|\rho(a),\nonumber\\
&&y_{j}^{q_{0j}} = y_{i}^{-\frac{1}{2} \sum_\Phi \sum_{\rho\in R_\Phi} |\rho(m)| F_i (\Phi)}, \nonumber \\
&&\epsilon_0 = \frac{1}{2} \sum_\Phi (1-\Delta_\Phi) \sum_{\rho\in
R_\Phi} |\rho(m)|
- \frac{1}{2} \sum_{\alpha \in G} |\alpha(m)|, \nonumber\\
&&f_{chiral}\left(e^{ia}, y_{j},x\right) = \sum_\Phi \sum_{\rho\in R_\Phi}
\left[ e^{i\rho(a)} y_{j}^{F_{j}}
\frac{x^{|\rho(m)|+\Delta_\Phi}}{1-x^2}  -  e^{-i\rho(a)}
y_{j}^{-F_{j}} \frac{x^{|\rho(m)|+2-\Delta_\Phi}}{1-x^2}
\right],\nonumber\\
&&f_{vector}\left(e^{ia},x\right)=-\sum_{\alpha\in G}e^{i\alpha(a)}x^{|\alpha(m)|}
\label{universal}
\end{eqnarray}
where $\sum_\Phi$, $\sum_{\rho\in R_\Phi}$ and $\sum_{\alpha\in G}$
represent the summations over all chiral multiplets, all weights and
all roots, respectively. $F_i$ are the Cartan generators acting only
on the $i$-th flavor.

The index formula for the duality that we are considering is similar
to that for the Giveon-Kutasov duality \cite{HKPP11} except for the
absence of the CS term and the contribution of the additional gauge
singlet chiral multiplet $Y$ on the magnetic side.
 It's important to take account
of nontrivial action of $Z_2$ element in $O(N)$ whose determinant is
$-1$. We review it following \cite{HKPP11}. Let us first consider
$O(2N)$ case.  Since the weights of the fundamental representation
are $\pm \epsilon_i$ where $i=1,\cdots,N$ and  the roots of $O(2N)$
are $\pm\epsilon_i\pm\epsilon_j$ where $i,j=1,\cdots,N$ and $i\neq
j$, we obtain for example the contribution from one chiral multiplet
with vector representation with turning off the chemical potential
$y_j=1$.
\begin{equation}
f_{chiral}(e^{ina}, y_j=1,x^n) =
\frac{ x^{nr} -  x^{(2-r)n}}{1-x^{2n}}   \left[ \sum^{N}_{i=1}  x^{n|m_i|} 2\cos n a_i  \right]
\end{equation}
This  formula holds for $SO(2N)$ case. We should consider the
additional projection for $Z_2$ element of $O(2N)$ not belonging to
$SO(2N)$ group. We choose the
specific $Z_2$ action,
\begin{equation}
Z_2= \left(\begin{array}{cccc}
1&&&\\
&-1&&\\
&&1&\\
&&&\ddots
\end{array}\right).
\end{equation}
Under this $Z_2$ action, the eigenvalues of the holonomy and the
monopoles are mapped to
\begin{equation}
e^{\pm ia_1}\rightarrow\pm1,~~~~~~\pm m_1\rightarrow0.
\end{equation}
The other variables are not affected. Thus, $f_{chiral}$ turns into
\begin{equation}
f_{chiral}(e^{ina}, y_j=1,x^n) =
 \frac{ x^{nr} -  x^{(2-r)n}}{1-x^{2n}}   \left[ (1+(-1)^n) + \sum^{N}_{i=2} x^{n|m_i|} 2\cos n a_i  \right],
\end{equation}

Let us turn to $O(2N+1)$ theory. The weights of the fundamental
representation are $\pm \epsilon_i$ where $i=1,\cdots,N$ and  the
roots of $O(2N+1)$ are $\pm \epsilon_i$ and
$\pm\epsilon_i\pm\epsilon_j$ where $i,j=1,\cdots,N$ and $i\neq j$.
In this case, we choose $Z_2$ action,
\begin{equation}
Z_2= \left(\begin{array}{cccc}
1&&&\\
&\ddots&&\\
&&1&\\
&&&-1
\end{array}\right),
\end{equation}
where an eigenvalue 1 of the holonomy in the fundamental
representation is mapped to
\begin{equation}
1\rightarrow-1
\end{equation}
while the others are not influenced. Furthermore, eigenvalues
$e^{\pm ia_i}$ of the holonomy in the adjoint representation are
transformed to
\begin{equation}
e^{\pm ia_i}=e^{\pm ia_i}\cdot1\rightarrow e^{\pm ia_1}\cdot(-1)
\end{equation}
while the others, which are in the form of $e^{i(\pm a_i\pm
a_j)}=e^{\pm ia_i}\cdot e^{\pm ia_i}$  are not influenced.

 The result of the index computation is
given in Table \ref{index}. The indices on both sides agree perfectly.
\begin{longtable}{|c|c|c|p{7.5cm}|}
\hline
  &  Electric  &  Magnetic  &  \\
$(N_f,N_c)$  &  $O(N_c)$  &  $O(N_f - N_c + 2)$  &  Index ($r$ is the $R$-charge.)\\
\hline
(0,1)       &   $O(1)$   &  $O(1)$         &  $1$\\
\hline
(1,1)       &   $O(1)$   &  $O(2)$         &  $1-x^2-2 x^4-2 x^6+x^{6-2 r}+x^{2 r}+x^{4 r}+x^{6 r}+x^{8 r}+\cdots$\\
\hline
(2,1)       &   $O(1)$   &  $O(3)$         &  $1-4 x^2-5 x^4+x^{4-2 r}+7 x^{6 r}+x^{2 r} \left(3-4 x^2\right)+x^{4 r} \left(5-4 x^2\right)+\cdots$\\
\hline
(3,1)       &   $O(1)$   &  $O(4)$         &  $1-9 x^2+3 x^{4-2 r}+28 x^{6 r}+x^{4 r} \left(15-33 x^2\right)+x^{2 r} \left(6-21 x^2\right)+\cdots$\\
\hline
(1,2)       &   $O(2)$   &  $O(1)$         &  $1-x^2-2 x^4+x^{5-5 r}+x^{4-4 r}+x^{3-3 r}+x^{2-2 r}+x^{1-r}+x^{2 r}+x^{4 r}+x^{3+r}+\cdots$\\
\hline
(2,2)       &   $O(2)$   &  $O(2)$         &  $1-4 x^2+5 x^4+x^{4-4 r}+6 x^{4 r}+x^{2 r} \left(3-8 x^2\right)+x^{-2 r} \left(x^2+x^4\right)+\cdots$\\
\hline
(3,2)       &   $O(2)$   &  $O(3)$         &  $1-9 x^2+36 x^4+x^{3-3 r}+3 x^{4-2 r}+21 x^{4 r}+x^{2 r} \left(6-45 x^2\right)+\cdots$\\
\hline
(4,2)       &   $O(2)$   &  $O(4)$         &  $1-16 x^2+148 x^4+x^{4-4 r}+6 x^{4-2 r}+55 x^{4 r}+x^{2 r} \left(10-144 x^2\right)+\cdots$\\
\hline
(2,3)       &   $O(3)$   &  $O(1)$         &  $1+3 x+x^2+x^{4-8 r}+x^{3-6 r}+3 x^{2 r}+x^{-2 r} \left(x+3 x^2\right)+x^{-4 r} \left(x^2+3 x^3\right)+\cdots$\\
\hline
(3,3)       &   $O(3)$   &  $O(2)$         &  $1-9 x^2+x^{4-6 r}+6 x^{4-4 r}+x^{2-3 r}+6 x^{2-r}+6 x^{2 r}+21 x^{4 r}+15 x^{2+r}+\cdots$\\
\hline
(4,3)       &   $O(3)$   &  $O(3)$         &  $1-16 x^2+35 x^3+x^{3-4 r}+10 x^{3-2 r}+10 x^{2 r}+55 x^{4 r}+\cdots$\\
\hline
(5,3)       &   $O(3)$   &  $O(4)$         &  $1-25 x^2+x^{4-5 r}+15 x^{4-3 r}+10 x^{4-2 r}+120 x^{4 r}+x^{2 r} \left(15-350 x^2\right)+\cdots$\\
\hline
(3,4)       &   $O(4)$   &  $O(1)$         &  $1+46 x^2+21 x^{4-8 r}+21 x^{3-5 r}+6 x^{2-4 r}+x^{1-3 r}+21 x^{2-2 r}+6 x^{1-r}+6 x^{2 r}+21 x^{1+r}+\cdots$\\
\hline
(4,4)       &   $O(4)$   &  $O(2)$         &  $1+39 x^2+x^{4-8 r}+10 x^{4-6 r}+x^{2-4 r}+10 x^{2-2 r}+10 x^{2 r}+\cdots$\\
\hline
(5,4)       &   $O(4)$   &  $O(3)$         &  $1-25 x^2+x^{3-5 r}+15 x^{3-3 r}+120 x^{3-r}+15 x^{2 r}+120 x^{4 r}+\cdots$\\
\hline
(6,4)       &  $O(4)$    &   $O(4)$        &  $1-36 x^2+x^{8-12 r}+21 x^{4-4 r}+246 x^{4-2 r}+231 x^{4 r}+ 21 x^{2 r}-720 x^{2+2r}+x^{4-6 r}+\cdots$\\
\hline
\caption{The result of the superconformal index computation.}\label{index}
\end{longtable}

We can examine the chiral ring structure of the theory. Let us first
consider the $N_c>1$ cases. In these cases, the chiral primaries on
the electric side are the meson operators $M^{\{ab\}}=Q^a_iQ^b_i$
and the monopole operator $Y$. One can show by  following
closely\cite{Bashkirov}, there are no other monopole operators which
are chiral primary. The baryon operators are projected out due to
the nontrivial $Z_2$ element of $O(N)$ whose determinant is $-1$.
Their counterparts on the magnetic side are the lowest components of
the gauge singlet chiral multiplets $M^{\{ab\}}$ and $Y$. The
composite meson operators $m_{\{ab\}}=q^i_aq^i_b$ and the monopole
operator $y$ are all $Q$-exact due to the superpotential
\eqref{superpotential}. Thus, the chiral ring structures on both
side are exactly the same.  Their contribution to the superconformal
index can be easily checked. There are $\frac{N_f(N_f+1)}{2}$ meson
operators $M^{\{ab\}}$ of energy $\epsilon=2r$, whose contributions
to the index is thus $\frac{N_f(N_f+1)}{2}x^{2r}$. The monopole
operator $Y$, which has energy $\epsilon=N_f-N_c+2-N_f r$, makes the
contribution $x^{N_f-N_c+2-N_f r}$ to the index.

For $N_c=1$ and $N_f-N_c+2\neq1$, on the other hand, there is no
monopole operator on the electric side because the gauge group is
just $O(1)=\mathbb Z_2$. Thus,  the meson operators $M^{\{ab\}}$ are
the only chiral primaries. On the magnetic side, there is still the
chiral operator $Y$, which seems to be one of the chiral primaries.
We propose that it becomes $Q$-exact due to the superpotential
\begin{equation}
W\sim Yy+ y^2{\det m}\sim\frac{Y^2}{\det m}.
\label{specialpotential}
\end{equation}
The form of the superpotential is similar to the Afflek-Dine-Seiberg (ADS)
superpotential \cite{ADS}. Here simple $R$ charge counting shows that such superpotential is possible due
to the additional singlet
$Y$ to soak up the additional fermion zero modes. It would be
interesting to derive this by explicit computation. In the index
computation,  there is no $x^{N_f+1-N_f r}$ term in the index
because the contribution $-x^{N_f+1-N_f r}$ of the fermionic
operator $\psi^\dagger_Y\det m$ exactly cancels out the contribution
$x^{N_f+1-N_f r}$ of $Y$ as we expected. Indeed, the contribution of
the fermionic operators $m^{N_f}\psi^\dagger_Y$ is in general
canceled by the contribution of the scalar operators $m^{N_f}y$.
Here the contracted gauge indices of
$m^{N_f}y=(q^i_aq^i_b)^{N_f}y$ run over $N_f-1=N_c'-2$ values corresponding
to the unbroken gauge group $O(N_c'-2)$ in the presence of the monopole flux
associated with  $y$.
Here $N_c'$ denotes the magnetic gauge group.
Schematically we have the block-diagonal structure
\begin{equation}
\left( \begin{array}{cc}
                m & 0 \\
               0 & y      \end{array}  \right).
\end{equation}
The mesons
$m_{\{ab\}}$ do not couple to the magnetic flux excited by $y$ and
remain as scalars. It is obvious that Gauss constraint is satisfied if we view such
operator as a state defined on $S^2 \times R$. However, one combination of such scalar
operators, $y\det m$, is vanishing since $N_f\times N_f$ matrix $m$ has rank $N_f-1$.
 As a result, $\psi^\dagger_Y\det m$ can
survive and cancel the contribution of $Y$. Therefore,  meson
operators $M^{\{ab\}}$ are the only chiral primaries for $N_c=1$ on
either side.

For $N_c=1$ and $N_f-N_c+2=1$, we have $N_f=0$ so that both sides
have abelian gauge group.  In this case, there would be no monopole
operators, no meson operators and the theory is trivial. Thus, the
superconformal index is just 1 on either sides.

Furthermore, we can trace not only the contributions of the chiral
primaries but also those of BPS operators with nonzero angular
momentum. For example, there are ${N_f}^2$ fermionic operators
$Q^a_i\psi^{\dagger i}_b$ on the electric side of energy
$\epsilon=\frac{3}{2}$ and the angular momentum $j=\frac{1}{2}$.
Correspondingly, there are $N_f^2$ fermionic operators
$q^i_a\psi^{\dagger b}_{q i}$ on the magnetic side of same energy
and the same angular momentum. On either side, their contribution to
the index
 is $-{N_f}^2x^2$. The results listed in Table \ref{index}, except for $(N_f,N_c)=(2,3)$, $(3,4)$ and $(4,4)$ cases,
 confirm this argument. The three exceptional cases are due to the fact that there are  additional BPS
 operators whose $\epsilon+j$ is 2. Let us examine these exceptional cases in detail.

At first, we consider the $(N_f,N_c)=(2,3)$ case. On the electric
side, we can find the operators $M^2Y^2$, which are scalar BPS
operators, of energy $\epsilon=2$. The contracted gauge indices of
$M^2Y^2=Q^a_iQ^b_iQ^c_jQ^d_jY^2$ run over the values corresponding
to the unbroken gauge group in the presence of monopole flux
associated with $Y$ such that $M^2$ do not couple to the magnetic
flux. If the operators $M^2$ couple to the magnetic flux, then they
get an effective spin such that their energy are no longer 2. Since
the unbroken gauge group is in this case just $O(1)$, the gauge
indices are fixed to the one value corresponding to the unbroken
$O(1)$. Thus, the operators $M^2Y^2=Q^a_iQ^b_iQ^c_iQ^d_iY^2$ with
the fixed gauge index $i$ make the contribution ${}_2H_4x^2=5x^2$ to
the
index.\footnote{${}_nH_m={}_{n+m-1}C_m=\frac{(n+m-1)!}{m!(n-1)!}$ is
the combination with repetition.} Therefore, the total $x^2$ term is
$(-4+5)x^2=x^2$. Here $-4$ comes from $-N_fx^2$ discussed in the
previous paragraph. On the magnetic side, as opposed to the electric
side, there is no gauge index to be contracted and no issue of the
coupling to the flux because $M^{\{ab\}}$ and $Y$ are just
elementary chiral fields. Thus, the number of $M^2Y^2$ is
$3\cdot4/2=6$ where we have $2\cdot3/2=3$ $M^{\{ab\}}$. However,
some of its contribution is canceled by that of the fermionic
operator $Y\psi^\dagger_Y$, whose contribution is $-x^2$. Therefore,
the total $x^2$ term in the index is again $(-4+6-1)x^2=x^2$ on the
magnetic side.

The $(N_f,N_c)=(3,4)$ case is exactly the same. On the electric
side, the scalar BPS operators
$M^3Y^2=Q^a_iQ^b_iQ^c_jQ^d_jQ^e_kQ^f_kY^2$ have energy $\epsilon=2$
where the contracted gauge indices $i$, $j$ and $k$ run over 2
values corresponding to the unbroken gauge group $O(2)$. Since the
number
 of $M^{\{ab\}}$ is $3\cdot4/2=6$, there are naively ${}_6H_3=56$ $M^3Y^2$. However, we could check that
 $\det M=0$ if the gauge indices run over only 2 values, which means that the number of independent $M^3Y^2$
 is $56-1=55$. On the magnetic side, there are 56 $M^3Y^2$ and one $Y\psi^\dagger_Y$. They make an additional
 contribution $(56-1)x^2=55x^2$. Thus, on either electric and magnetic side, the total $x^2$ term is $(-9+55)x^2=46x^2$.

 For $(N_f,N_c)=(4,4)$, the scalar BPS operators $M^2Y=Q^a_iQ^b_iQ^c_jQ^d_jY$ on the electric side have energy
 $\epsilon=2$ where the contracted gauge indices $i$ and $j$ run  over 2 values corresponding to the unbroken
 $O(2)$ as before. Since there are $4\cdot5/2=10$ $M^{\{ab\}}$ and ${}_{10}H_2=55$ $M^2$, the contribution of
 $M^2Y$ to the index is $55x^2$. On the magnetic side, there are 55 $M^2Y$ as on the electric side. Therefore, the
 total $x^2$ term is $(-16+55)x^2=39x^2$ on either sides.

We can check the indices in more detail by turning on the chemical
potentials of the global flavor symmetry. There are few differences
in index formula that there are flavor charge terms $y_j^{q_{0j}}$
and chemical potentials $y_j$ are in letter index
\cite{Imamura11,HKPP11}. The resulting indices can be rewritten in
terms of $U(1)\times SU(N_f)$ characters. All of the above results
correspond to the cases with setting $y_j=1$. For example, for the
case of $N_c=2$, $N_f=2$, either indices of electric and magnetic
theory is
\begin{eqnarray}
I&=&1-2x^{2-2r}\frac{1}{y_1 y_2}-x^2\left(\frac{y_1}{y_2}+\frac{y_2}{y_1}+2\right)+x^{2r}\left(y_1^2+y_1 y_2+y_2^2\right)
+\cdots\nonumber\\
&=&1-2x^{2-2r}y_0^{-2}-x^2\left(\chi_{1}\left(u\right)+1\right)+x^{2r}y_0^2\chi_{1}\left(u\right)+\cdots
\end{eqnarray}
where $y_0=(y_1 y_2)^{1/2}$ and $u=y_1/y_2$ correspond to the
chemical potentials for the global symmetry $U(1)_A\times SU(2)$ and
 $\chi_n(u)=u^{-n}+u^{-n+1}+\ldots+u^n$ are the characters of
$SU(2)$.

As a final remark, Seiberg-like duality for the orthogonal gauge
group considered in \cite{Kapustin11,HKPP11} can be derived from the
duality we have considered here. It is well known that the CS term
$\pm\frac{1}{8\pi}\int {\rm tr}A\wedge F$, which is one unit of the
CS level for the orthogonal groups, is generated by integrating out
a charged fermion by giving it  axial mass. This mass term can be
understood as arising from weakly gauging the axial symmetry
$U(1)_A$ by a background vector field $V_{mass}=-i\theta
\bar{\theta} \mu$. Thus the mass term for a chiral multiplet $Q$ on
the electric side with $U(1)A$ charge $+1$ is given by
\begin{equation}
 {\cal L}_{mass}=\int d^4\theta Q^{\dagger}e^{V_{mass}}Q.
\end{equation}
Note that on the magnetic theory under Aharony duality, $q$ picks up
axial mass term of negative sign since it has $U(1)_A$ charge $-1$.
Thus on the electric side, we flow from $O(N_c)$ with $N_f$ flavors into
$O(N_c)$ with $N_f-1$ flavors with Chern-Simons level 1 while in the
magnetic side, we are flowing to $O(N_f-N_c+2)$ with $N_f-1$ flavors
with Chern-Simons level $-1$. Thus from Aharony duality, one can
obtain Seiberg-like duality for Chern-Simons matter theories. By
repeatedly integrating out charged chiral multiplets one can obtain
Seiberg-like duality for Chern-Simons matter theories with higher
Chern-Simons level. Note also that by weakly gauging $U(1)_A$
symmetry, one also gives mass to the monopole operators $Y,y$.

\section{Conclusions}
In this paper we provide evidences for Aharony duality for orthogonal
groups. By using available tools of the partition function and the superconformal
index, we give sufficient evidences for Aharony duality for orthogonal gauge groups with
matters in the vector representation.
Along the investigation, we come up with the proposal that for $O(N_f+1)_{N_f}$ theory
in the magnetic side, it should develop the superpotential
\begin{equation}
W=\frac{Y^2}{\det m}
\end{equation}
reminiscent of the ADS superpotential in 4d. Recall that without the
additional singlet $Y$ we could not have such superpotential so it
would be interesting to work out the proposed supertpotential
explicitly. Further we expect that for $N_f-N_c+2=0$ there would be
no SCFT dual in the magnetic side. Rather, we will have the quantum
moduli space to be modified
\begin{equation}
Y^2\det M=1.
\end{equation}
This is another interesting exercise for instanton calculus.
Analogous 4d computation was done in \cite{WittenBeasley, ParkMatsuo}.

\vskip 0.5cm  \hspace*{-0.8cm} {\bf\large Acknowledgements} \vskip
0.2cm

\hspace*{-0.75cm} We are grateful to A. Kapustin for the discussion
on the Aharony duality. J.P.  is supported by the KOSEF Grant
R01-2008-000-20370-0, the National Research Foundation of Korea
(NRF) Grants No. 2009-0085995  and 2005-0049409 through the Center
for Quantum Spacetime (CQUeST) of Sogang University.
J. P. appreciates APCTP for its stimulating environment for
  research and acknowledges Simons summer workshop on mathematics and
  physics 2011 for hospitality while a part of the  current work was carried out .


\begin{thebibliography}{1000}

\bibitem{Schwarz04}
  J.~H.~Schwarz,
  ``Superconformal Chern-Simons theories,''
  JHEP {\bf 0411} (2004) 078
  {\tt [arXiv:hep-th/0411077]}.

\bibitem{BL1}
  J.~Bagger and N.~Lambert,
  ``Modeling multiple M2's,''
  Phys.\ Rev.\  D {\bf 75} (2007) 045020
  {\tt [arXiv:hep-th/0611108]}.

\bibitem{BL2}
  J.~Bagger and N.~Lambert,
  ``Gauge symmetry and supersymmetry of multiple M2-branes,''
  Phys.\ Rev.\  D {\bf 77} (2008) 065008
  {\tt [arXiv:0711.0955 [hep-th]]}.

\bibitem{BL3}
  J.~Bagger and N.~Lambert,
  ``Comments on multiple M2-branes,''
  JHEP {\bf 0802} (2008) 105
  {\tt [arXiv:0712.3738 [hep-th]]}.

\bibitem{gus1}
  A.~Gustavsson,
  ``Algebraic structures on parallel M2-branes,'' Nucl.\ Phys.\  B {\bf 811} (2009) 66
  {\tt [arXiv:0709.1260 [hep-th]]} .

\bibitem{gus2}
  A.~Gustavsson,
  ``Selfdual strings and loop space Nahm equations,'' JHEP {\bf 0804} (2008)
  083,  {\tt [arXiv:0802.3456 [hep-th]]}.

\bibitem{GaiottoWitten}
  D.~Gaiotto and E.~Witten,
  ``Janus configurations, Chern-Simons couplings, and the theta-angle in N=4 Super Yang-Mills theory,''
  {\tt [arXiv:0804.2907 [hep-th]]}.

 \bibitem{Hosomichi08}
  K. Hosomichi,  K. Lee,  S. Lee ,S. Lee
    and J. Park,
  ``N=4 Superconformal Chern-Simons Theories with Hyper and Twisted Hyper
    Multiplets,''
  JHEP {\bf 0807} (2008) 091,
 {\tt
[arXiv:0805.3662 [hep-th]]}.

\bibitem{Aharony08}
 O.  Aharony, O. Bergman, D.  Jafferis and J.  Maldacena,
  ``N=6 superconformal Chern-Simons-matter theories, M2-branes and their
    gravity duals,''
  JHEP {\bf 0810} (2008) 091
  {\tt [arXiv: 0806.1218 [hep-th]]}.

\bibitem{Hosomichi08a}
K. Hosomichi, K. Lee, S. Lee, S. Lee
    and J. Park,
  ``N=5,6 Superconformal Chern-Simons Theories and M2-branes on
  Orbifolds,''
  JHEP {\bf 0809} (2008) 002,
  {\tt [arXiv:0806.4977 [hep-th]]}.

\bibitem{Imamura08}
Y. Imamura and K. Kimura, `` N=4 Chern-Simons theories with
auxiliary vector multiplets,'' JHEP {\bf 0810} (2008) 040, {\tt
[arXiv:0807.2144 [hep-th]]}.



 \bibitem{Gaiotto07}
  D. Gaiotto and X. Yin,
  ``Notes on Superconformal Chern-Simons-Matter Theories,''
  JHEP {\bf 0708} (2007) 056,
  {\tt [arXiv:0704.3740 [hep-th]]}.


\bibitem{Hanany1}
S. Franco , A. Hanany , J. Park and   D. Rodriguez-Gomez, ``Towards
M2-brane Theories for Generic Toric Singularities,'' JHEP {\bf 0812}
(2008) 110,  {\tt  [arXiv:0809.3237 [hep-th]]}.

\bibitem{Hanany2}  A. Hanany, D. Vegh and  A. Zaffaroni,
``Brane Tilings and M2 Branes,''  JHEP {\bf 0903} (2009) 012, {\tt
 [arXiv:0809.1440 [hep-th]]}.

\bibitem{Hanany3} A. Hanany,
and A. Zaffaroni,  ``Tilings, Chern-Simons Theories and M2 Branes,''
JHEP 0810 (2008) 111,  {\tt [arXiv:0808.1244 [hep-th]]}.

\bibitem{Martelli}
D. Martelli and J. Sparks, ''Notes on toric Sasaki-Einstein
seven-manifolds and AdS(4) /
 CFT(3),'' JHEP {\bf 0811} (2008) 016, {\tt  [arXiv:0808.0904
 [hep-th]]}.

\bibitem{Jafferis09}
D. Gaiotto and D. Jafferis, ``Notes on adding D6 branes wrapping
$RP^3$ in $AdS(4) x CP^3$, {\tt  [arXiv:0903.2175 [hep-th]]}.

\bibitem{Benini10}
 F. Benini , C. Closset and S. Cremonesi,`` Chiral flavors and
M2-branes at toric CY4 singularities,'' JHEP {\bf 1002} (2010) 036.
{\tt [arXiv:0911.4127 [hep-th]]}.

\bibitem{KWY1}
A. Kapustin, B. Willett, I. Yaakov, ''Exact Results for Wilson Loops in Superconformal Chern-Simons Theories with Matter,''
 JHEP {\bf1003} (2010) 089, {\tt [arXiv:0909.4559 [hep-th]]}.

\bibitem{Kapustin11}
  A. Kapustin,
  ``Seiberg-like duality in three dimensions for orthogonal gauge
  groups, ''
  {\tt [arXiv:1104.0466 [hep-th]]}.

\bibitem{Willett11}
  B. Willett and I. Yaakov,
  ``N=2 Dualities and Z Extremization in Three Dimensions,''
  {\tt [arXiv:1104.0487 [hep-th]]}.


\bibitem{Bhattacharya09}
  J. Bhattacharya and S. Minwalla,
  ``Superconformal Indices for $ N=6$ Chern Simons Theories,''
  JHEP, {\bf 0901} 2009) 014,
 {\tt [arXiv:0806.3251 [hep-th]]}.

\bibitem{Kim09}
  S. Kim,
  ``The complete superconformal index for N=6 Chern-Simons theory,''
  Nucl. Phys. {\bf B821} (2009) 241,
  {\tt [arXiv:0903.4172 [hep-th]]}.


\bibitem{Imamura11}
  Y. Imamura and S. Yokoyama,
  ``Index for three dimensional superconformal field theories with general R-charge assignments,''
  {\tt [arXiv:1101.0557 [hep-th]]}.

\bibitem{HKPP11}
  C. Hwang, H. Kim, K.-J. Park, J. Park,
  ``Index computation for 3d Chern-Simons matter theory: test of Seiberg-like duality,''
  {\tt [arXiv:1107.4942 [hep-th]]}.

\bibitem{Bashkirov} D. Bashkirov, ``Aharony duality and monopole operators in three
dimensions,'' {\tt [arXiv:1106.4110 [hep-th]]}.

  \bibitem{Giveon09}
  A. Giveon and D. Kutasov,
  ``Seiberg Duality in Chern-Simons Theory,''
  Nucl. Phys. {\bf B812} (2009) 1,
{\tt [arXiv:0808.0360 [hep-th]]}.

\bibitem{Niarchos} V. Niarchos, ``R-charges, Chiral Rings and RG Flows in Supersymmetric
Chern-Simons-Matter Theories,''
JHEP {\bf 0905} (2009) 054.

\bibitem{Koh11}

 D. Gang, E. Koh, K. Lee and J. Park,
``ABCD of 3d ${\cal N}=8$ and 4 Superconformal Field Theories,''
{\tt [arXiv:1108.3647 [hep-th]]}.

\bibitem{Vartanov11}

C. Krattenthaler, V.  Spiridonov, G.  Vartanov,  ``Superconformal
indices of three-dimensional theories related by mirror symmetry,''
JHEP {\bf 1106} (2011) 008, {\tt [arXiv:1103.4075 [hep-th]]}.




\bibitem{Aharony97}
 O. Aharony ``IR duality in d=3 $N$=2
  $Usp(2N_c)$ and $U(N_c)$ Gauge Theories''
 Nucl. Phys.{\bf  B404} (1997) 71,
 {\tt [arXiv:hep-th/9703215]}.

\bibitem{Jafferis10}
  D. Jafferis,
  ``The Exact Superconformal R-Symmetry Extremizes Z,''
{\tt [arXiv:1012.3210 [hep-th]]}.


\bibitem{Benini11}


 F. Benini, C. Closset, S. Cremonesi,
 ''Comments on 3d Seiberg-like dualities,''
 {\tt [arXiv:1108.5373 [hep-th]]}.

\bibitem{GiveonKutasov98}
 A. Giveon, D. Kutasov, ``Brane Dynamics and Gauge Theory,''
 Rev.Mod.Phys.{\bf 71} (1999) 983, {\tt [arXiv:hep-th/9802067]}.



\bibitem{Vartanov11-2}
 F. Dolan, V.  Spiridonov, G.  Vartanov,  ``From 4d superconformal
indices to 3d partition functions,'' Phys.Lett. {\bf B704} (2011)
234,  {\tt [arXiv:1104.1787 [hep-th]]}.

\bibitem{Vartanov11-3}
 V.  Spiridonov, G.  Vartanov, ``Elliptic hypergeometry of supersymmetric dualities
II. Orthogonal groups, knots, and vortices,'' {\tt [arXiv:1107.5788
[hep-th]]}.

\bibitem{Aharony11}
O. Aharony, I. Shamir, ``On $O(N_c)$ d=3 N=2 supersymmetric QCD
Theories,'' {\tt [arXiv:1109.5081 [hep-th]]}.

\bibitem{ADS}
I. Affleck, M, Dine and N. Seiberg, ``Dynamical supersymmmetry breaking in supersymmetric QCD,''
Nucl. Phys. {\bf B273} (1986) 629.





\bibitem{WittenBeasley}
 C. Beasley and  E. Witten, ``New Instanton Effects In Supersymmetric QCD,''
 JHEP {\bf 0501} (2005) 056,
{\tt [hep-th/0409149]}.

\bibitem{ParkMatsuo}

Y. Matsuo, J. Park, C. Ryou and M. Yamamoto,
``D-instanton derivation of multi-fermion F-terms in supersymmetric QCD,''
JHEP {\bf 0806} (2008) 051, {\tt [arXiv:0803.0798 [hep-th]]}.


\end{thebibliography}
\end{document}